\documentclass[fleqn,10pt]{wlscirep}
\usepackage[utf8]{inputenc}
\usepackage[T1]{fontenc}
\usepackage[load-configurations=abbreviations, redefine-symbols=true]{siunitx}
\usepackage{upgreek}   
\usepackage{hyperref}
\usepackage{graphicx}  
\usepackage{bm}        
\usepackage{verbatim}   
\usepackage{hyperref}
\usepackage{braket}
\usepackage{lineno}
\usepackage{textcomp}
\usepackage{xcolor}

\title{A pyramid MOT with integrated optical cavities as a cold atom platform for an optical lattice clock}

\author[1,2,*,+]{William Bowden}
\author[1,+]{Richard Hobson}
\author[1]{Ian ~R. Hill}
\author[1,3]{Alvise Vianello}
\author[1]{Marco Schioppo}
\author[1]{Alissa~Silva}
\author[1,2]{Helen~S.~Margolis}
\author[2]{Patrick~E.~G.~Baird}
\author[1,2,3]{Patrick Gill}

\affil[1]{National Physical Laboratory, Hampton Road, Teddington TW11 0LW, United Kingdom}
\affil[2]{Clarendon Laboratory, Parks Road, Oxford OX1 3PU, United Kingdom}
\affil[3]{Blackett Laboratory, Imperial College London, Prince Consort Road, London SW7 2AZ, United Kingdom}
\affil[*]{william.bowden@npl.co.uk}
\affil[+]{these authors contributed equally to this work}

\begin{abstract}
We realize a two-stage, hexagonal pyramid magneto-optical trap (MOT) with strontium, and demonstrate loading of cold atoms into cavity-enhanced 1D and 2D optical lattice traps, all within a single compact assembly of in-vacuum optics. We show that the device is suitable for high-performance quantum technologies, focusing especially on its intended application as a strontium optical lattice clock. We prepare \num{2e4} spin-polarized atoms of $^{87}$Sr in the optical lattice within \SI{500}{\milli\second}; we observe a vacuum-limited lifetime of atoms in the lattice of \SI{27}{\second}; and we measure a background DC electric field of \SI{12}{\volt\per\meter} from stray charges, corresponding to a fractional frequency shift of $(-1.2 \pm 0.8) \times 10^{-18}$ to the strontium clock transition. When used in combination with careful management of the blackbody radiation environment, the device shows potential as a platform for realizing a compact, robust, transportable optical lattice clock with systematic uncertainty at the \num{e-18} level.
\end{abstract}
\begin{document}

\flushbottom
\maketitle

\thispagestyle{empty}

\section*{Introduction}

Single beam pyramid magneto-optical traps (MOTs) are well-suited for cold-atom quantum technologies \cite{Arlt1998} as their compact, elegant design simplifies the optical setup needed for laser cooling. They have already been used to realise microwave clocks \cite{Xu2008}, cold atom sources \cite{Camposeo2001, kohel2003}, and inertial sensors---both lab-based \cite{Wu2017} and transportable \cite{Bodart2010} systems. However, a pyramid MOT has not previously been applied to cool alkaline-earth atoms for an optical lattice clock (OLC) \cite{Nemitz2016, LeTargat2013, Beloy2018, Campbell2017, Koller2017, Grotti2018}. Since atoms in an OLC must be cooled to the \si{\micro\kelvin} regime, there is a requirement in such systems to implement two MOTs at different wavelengths: a first stage on a strong cycling transition is used to capture atoms from a slow source \cite{Xu2003}, and then a second stage on a narrow line further cools and compresses the atomic cloud \cite{Katori1999, Mukaiyama2003}. These cooling transitions, along with the repumping and clock transitions, are shown in figure \ref{fig:levels}. Two-color pyramid MOTs have been previously demonstrated---initially as a dual-species cold atom source \cite{Harris2008} and more recently for the two-stage cooling of strontium \cite{he2017}---but have yet to be used to realize a functional OLC.


The two-stage MOT is insufficient by itself to realize an OLC; atoms must also be loaded into a magic-wavelength optical lattice trap so that the clock transition can be probed in the Lamb-Dicke regime \cite{Ye2008}. This motivates the construction of a single monolithic structure capable of supporting both a pyramid MOT and optics to generate the lattice trap. So far, devices which combine two different trapping methods have been limited to chip-based magnetic traps combined with either 1D \cite{Gallego2009} or 3D \cite{Straatsma2015} optical lattices. For both of these demonstrations, the lattice was formed of a single retro-reflected beam. In contrast, for OLCs it is advantageous to use a build-up cavity for the lattice \cite{Schiller2012, LeTargat2013, Akatsuka2010}. The enhancement provided by the cavity eliminates the need for high power lasers---a benefit for compact systems---and facilitates better characterization of lattice-induced sources of systematic uncertainty. To exploit these advantages, we integrate two build-up cavities into the same structure as the pyramid MOT which can either generate independent 1D lattices or be combined to make a 2D lattice. The benefit of the latter configuration is that the tight confinement provided by the 2D trap helps to suppress collisional effects which can perturb the clock frequency and lead to decoherence \cite{swallows2011}. 


Designing a highly integrated device capable of both laser cooling and trapping strontium is not without its challenges. This is especially true when realizing an accurate frequency standard as the environment surrounding the atoms must be well controlled. For example, the accumulation of charge on dielectric materials, e.g. cavity or MOT mirrors, positioned close to the atoms can result in significant electric fields which induce a Stark shift of the clock transition \cite{Beloy2018, Lodewyck2012}. Another important consideration when using an enhancement cavity is suppressing excessive atomic heating caused by instabilities in the intra-cavity intensity. In this report we outline how these challenges can be addressed in a compact cold atom platform, realizing the benefits of in-vacuum trapping optics without compromising the accuracy of the strontium optical lattice clock.

\section*{Results}

\subsection*{Opto-mechanical Design}
\label{sec:design}

\begin{figure}
    \centering
    \includegraphics[width=.5\columnwidth]{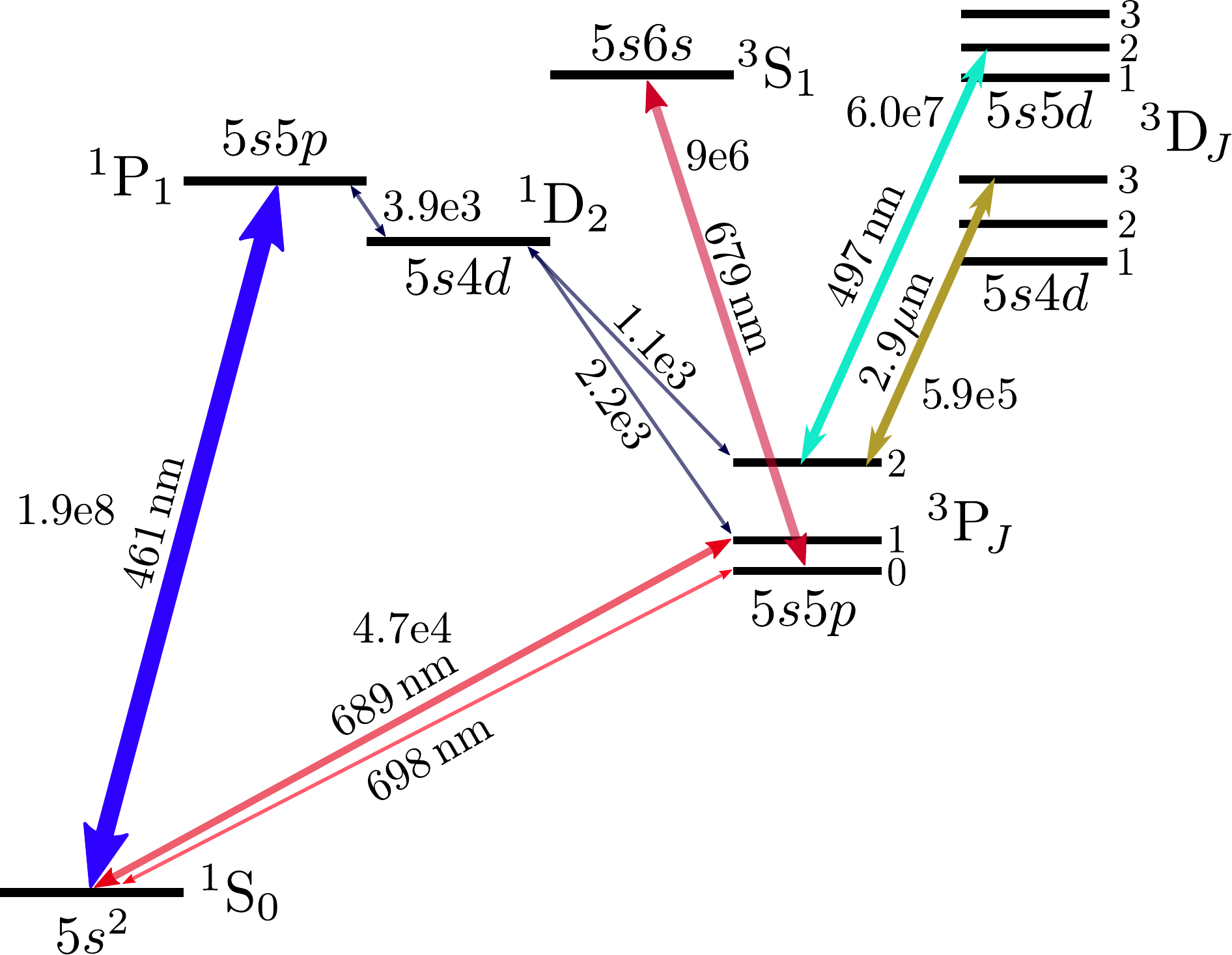}
    \caption{\label{fig:levels} Partial term diagram for strontium showing wavelengths and scatter rates (in units of s$^{-1}$) of the relevant transitions needed for laser cooling and precision spectroscopy. For clarity, hyperfine structure has been omitted.}
\end{figure}

\begin{figure}[tb]
    \centering
  \includegraphics[width=.95\columnwidth]{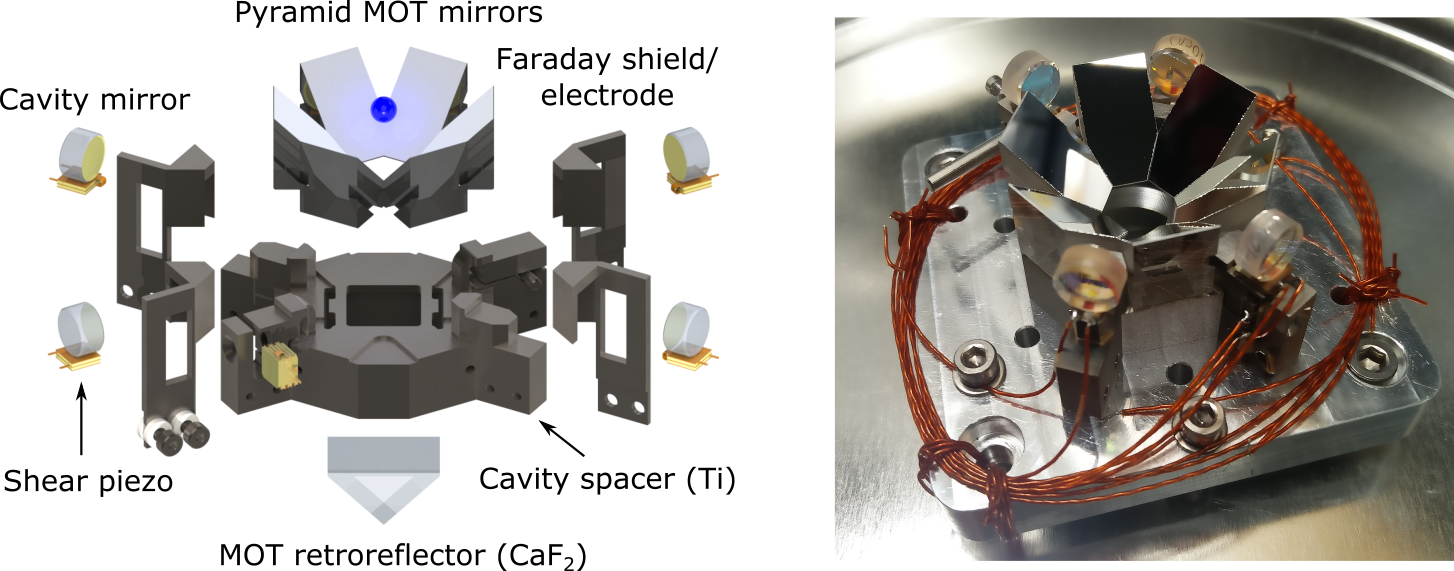}
    \caption{\label{fig:rendering} \textit{Left:} Exploded rendering of in-vacuum pyramid MOT and enhancement cavity setup. \textit{Right:} Photo of the partially completed assembly, before the installation of titanium electrodes. The two opposing pairs of cavity mirrors form two separate cavities rotated by 60 degrees; one cavity can be used by itself to realize a 1D optical lattice clock, and the other can optionally be used to create a 2D lattice. The third axis of optical access, rotated by 60 degrees from the cavities but in the same (\lq\lq radial\rq\rq ) plane, is where the Sr atomic beam enters the MOT region.}
\end{figure}

The pyramid MOT and cavity assembly is shown in figure \ref{fig:rendering}. A MOT is created by illuminating the assembly with a single, large (1/$e^2$ waist of \SI{22}{\milli\meter}), circularly polarized beam. Following the two-stage MOT, atoms are loaded into a 2D optical lattice trap that is generated by coupling a milliwatt of light at \SI{813}{\nano\meter} into the TEM$_{00}$ modes of the two-crossed enhancement cavities. A characterization of these traps using atomic data will be presented in following subsections. Here, we highlight some important aspects of the opto-mechanical design. More details regarding the construction of the cold atom platform are outlined in the Methods.

The pyramid MOT reflector consists of six silver-coated BK7 glass prisms and a CaF$_2$ right angle prism which are glued on to a titanium baseplate. Each prism is tapered in order to self-align to form a compact, hexagonal retro-reflector with an inradius measuring \SI{16.5}{\milli\meter}. The novel hexagonal symmetry is chosen partly to allow three radial axes of optical access and partly to produce an extra pair of radial confinement beams to compensate for the $\sim 40\%$ lower MOT beam intensity incident on the prism mirrors compared with the intensity at the center of the Gaussian beam. 

Incorporated into the design are two 1D lattice enhancement cavities which cross each other \SI{8}{\milli\meter} above the center of the CaF$_2$ retro-reflector to form a 2D lattice. The mirrors have a radius of curvature of \SI{10}{\centi\meter} and are separated by \SI{36}{\milli\meter}, resulting in a waist of \SI{100}{\micro\meter} for the magic wavelength lattice at \SI{813}{\nano\meter}. The two cavities both have a finesse of 6,000 at \SI{813}{\nano\meter}, providing an approximate power enhancement factor of 2,000. To probe the clock transition along the cavity axis, the \lq spectroscopy\rq\, cavity has an anti-reflection coating at \SI{698}{\nano\meter}. Additionally, this axis has a finesse of 13,000 at \SI{461}{\nano\meter}, enabling future investigation into cavity-enhanced, non-destructive detection of atoms using the $5s^2~^1\mathrm{S}_0 \rightarrow 5s5p~^1\mathrm{P}_1$ transition \cite{Vallet2017}.

Each mirror is glued on top of a piezoelectric actuator stack to provide up to \SI{6}{\micro\meter} tuning range of the length of each cavity. The actuator stacks consist of two $5 \times 5 \times \SI{0.5}{\milli\meter\cubed}$ shear piezos which are electrically contacted using squares of non-magnetic, ultra-high-vacuum (UHV) compatible copper-beryllium (CuBe) foil. The CuBe electrode in the middle of the stack is driven with voltages up to $\pm \SI{250}{\volt}$, but the electrodes at the top and bottom of the stack are grounded so that the inner high-voltage electrode is well shielded from the atoms. To provide further screening from the high voltages applied to the piezos, electrically isolated titanium shields are installed around the cavity mirrors. The titanium shields are then connected to separate pins of a vacuum feedthrough so that voltages can be applied to characterize any residual background electric fields. Details of this background field characterization are presented in later subsections.

To realize a 2D lattice, the modes of the two cavities must be well overlapped spatially. Unfortunately, the variation in mirror centration specified by the manufacturer precludes machining the entire assembly to the required precision to guarantee that the modes intersect. Instead, flexure mounts are machined into the titanium baseplate to enable kinematic adjustment of the mirrors in the non-spectroscopy cavity after gluing (see figure \ref{fig:rendering}). The flexure mounts provide adjustment of the tilt of each cavity mirror, thereby allowing us to move the non-spectroscopy cavity mode up or down the MOT axis until it intersects with the spectroscopy cavity. One flexure is adjusted by varying the diameter of the titanium dowel clamped underneath the flexure. This provides coarse adjustment of the cavity position with a resolution of around \SI{45}{\micro\meter}, and is initially aligned and tested outside of vacuum. The tilt angle of the other flexure was designed to be tunable using a shear stack piezo, which was expected to provide fine, bi-directional adjustment of the cavity mode position by $\pm \SI{50}{\micro\meter}$ after the assembly was put under vacuum. However, when this piezo was tested in vacuum, it was observed that the cavity mode did not move---possibly due to piezo failure or the stack coming unglued from the former. 
Fortunately, the initial out-of-vacuum alignment proved to be sufficient to generate a 2D lattice.

\subsection*{The Pyramid MOT}
\label{sec:pyramid_MOT}

\begin{figure}[tb]
    \centering
    \includegraphics[width=.85\columnwidth]{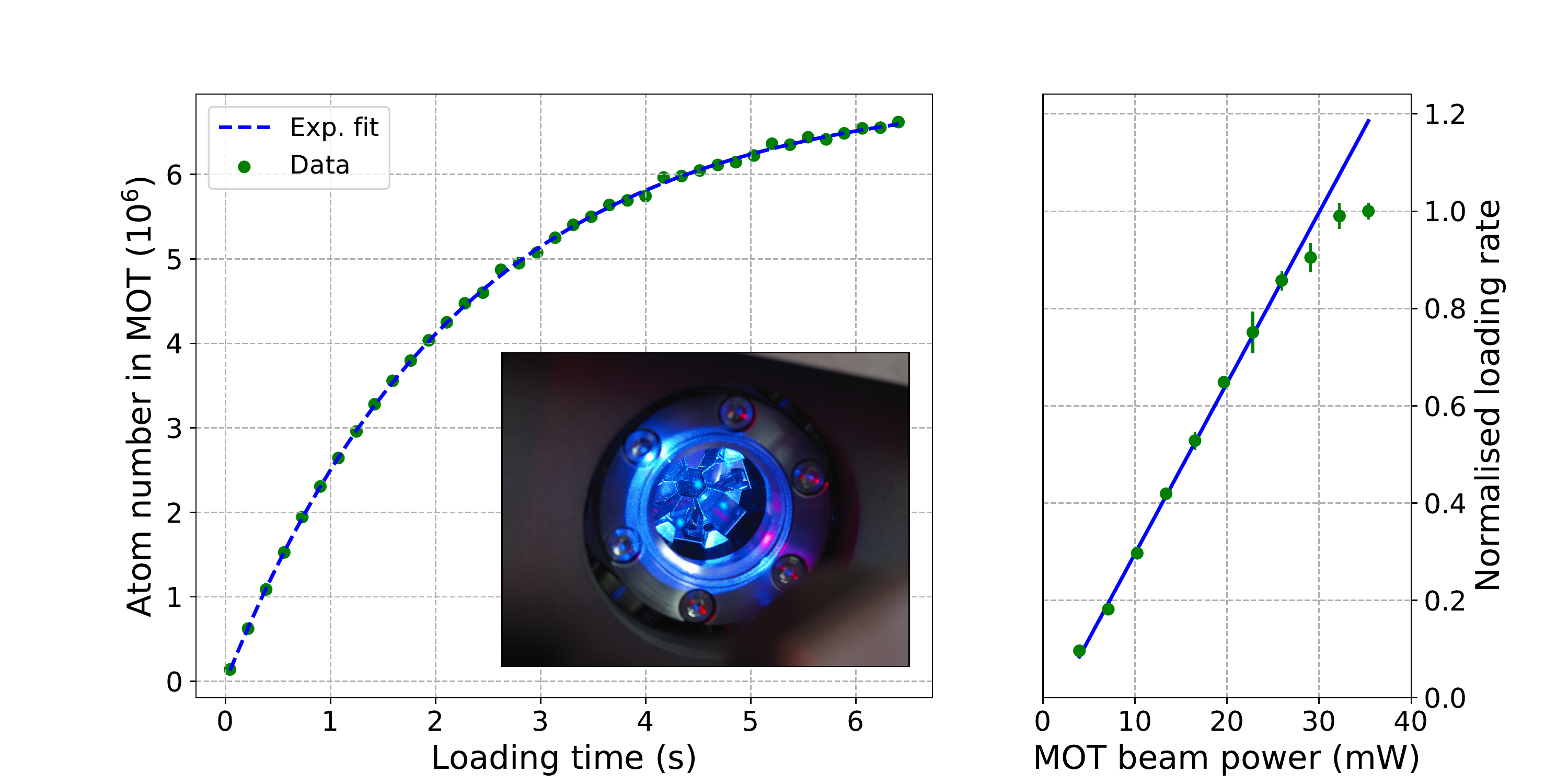}
    \caption{\label{fig:BlueMot} \textit{Left:} Loading curve for the $^{87}$Sr blue MOT, with an exponential fit revealing a loading rate of \SI{5e6}{atoms\per\second} and lifetime of \SI{2.3}{\second}. \textit{Inset:} Photograph of the $^{88}$Sr blue MOT. \textit{Right:} Dependence of $^{87}$Sr loading rate on MOT beam power.}
\end{figure}

Initial cooling and trapping is provided by a MOT operating on the $5s^2\,^1\mathrm{S}_0 \rightarrow 5s5p\,^1\mathrm{P}_1$ transition. The MOT is loaded from a slowed atomic beam generated by an effusive oven followed by a \SI{20}{\centi\meter} transverse-field permanent magnet Zeeman slower\cite{Hill2014}. Combined, the Zeeman slower and the MOT require \SI{100}{\milli\watt} of \SI{461}{\nano\meter} light which is generated by frequency doubling the output of a tapered amplifier at \SI{922}{\nano\meter} using a waveguide-enhanced second harmonic generation (SHG) module. After propagation, there is up to \SI{35}{\milli\watt} of available power in the slowing beam and \SI{40}{\milli\watt} in the MOT. Before hitting the pyramid MOT, the MOT beam is collimated to a 1/e$^2$ waist of \SI{22}{\milli\meter} so that each radial beam has approximately 60\% of the peak intensity at the axial center. This ensures that the cooling forces are roughly balanced between the axial and radial directions. At maximum MOT and slowing beam powers, the loading rate of $^{87}$Sr is \SI{5e6}{atoms\per\second} (see figure \ref{fig:BlueMot}). A faster loading rate is possible by increasing the oven temperature, but comes at the expense of having to refill the atomic source more frequently. 

The \SI{461}{\nano\meter} cooling transition is not fully closed. Approximately one in every \num{5e4} scatter events will result in an atom decaying to $5s4d\,^1\mathrm{D}_2$, from which 1/3 of the atoms decay to the metastable $5s5p\,^3\mathrm{P}_2$ manifold. In order to recycle these atoms back into the MOT, they are repumped by driving a transition to the $5s5d~^3\mathrm{D}_2$ state at \SI{497}{\nano\meter} which decays to the ground state via $5s5p\,^3\mathrm{P}_1$ \cite{Stellmer2014}. We use a waveguide enhanced SHG module to generate the repump light by frequency doubling an extended cavity diode laser operating at \SI{994}{\nano\meter}. The hyperfine structure arising from the presence of nuclear spin ($I = 9/2$) complicates repumping of $^{87}$Sr, since atoms can be shelved in any one of five different hyperfine states separated by a few \si{\giga\hertz} in energy. In our system, transitions from all five hyperfine states are driven by applying phase-modulation sidebands at \SI{710}{\mega\hertz} to the repump light using a waveguide electro-optical modulator (EOM) placed before the SHG module. The laser frequency is tuned to drive the $F = 11/2\rightarrow F'=11/2$ transition, while the other hyperfine states are repumped using the 1st- and 2nd order sidebands. To compensate any residual detuning between the hyperfine transitions and the sidebands, the laser linewidth is artificially broadened to \SI{50}{\mega\hertz} using the EOM. A weak decay path from $5s5d\,^3\mathrm{D}_2$ down to $5s5p\,^3\mathrm{P}_0$ necessitates an additional repump laser at \SI{679}{\nano\meter} which drives the $5s5p\,^3\mathrm{P}_0\rightarrow6s5p\,^3\mathrm{S}_1$ transition. With \SI{1}{\milli\watt} of power in each repump beam incident on the MOT, we measure a lifetime of \SI{2.3}{\second} based on a fit to the loading curve data shown in figure \ref{fig:BlueMot}.

The second-stage MOT, lattice loading and state preparation are all implemented using the transition at \SI{2.92}{\micro\meter} from metastable $5s5p\,^3\mathrm{P}_2\rightarrow 5s4d\,^3\mathrm{D}_3$. This technique will be described and characterized in detail in a future publication---here we briefly summarize the main points. During the final \SI{80}{\milli\second} of the blue MOT, the modulation sidebands on the \SI{497}{\nano\meter} repump beam are switched off so that atoms shelved into the $5s5p\,^3\mathrm{P}_2~F = 13/2$ state are no longer recycled into the blue MOT. At the same time, an additional beam at \SI{2.92}{\micro\meter} with a waist of \SI{26}{\milli\meter} is delivered to the pyramid MOT, capturing the metastable atoms in a MOT operating on the $5s5p\,^3\mathrm{P}_2~F = 13/2 \rightarrow 5s5d\,^3\mathrm{D}_3~F = 15/2$ transition. The \SI{2.92}{\micro\meter} beam has a power of \SI{700}{\micro\watt}, is red-detuned by an average of \SI{5}{\mega\hertz}, and is frequency modulated to cover a spectrum of \SI{3}{\mega\hertz} peak-to-peak, thus compensating for the range of Zeeman shifts introduced by the magnetic field gradient of \SI{0.7}{\milli\tesla\per\centi\meter}. At the end of this stage, approximately one fifth of the atoms are transferred from the blue MOT into the metastable MOT. To compress the cloud of metastable atoms, an additional \SI{80}{\milli\second} broadband MOT at \SI{2.92}{\micro\meter} is applied during which the mean detuning is ramped to \SI{2}{\mega\hertz} and the magnetic field gradient is ramped to \SI{0.2}{\milli\tesla\per\centi\meter}. Next, to attain a temperature of \SI{6}{\micro\kelvin} and load atoms into the optical lattice trap, a final \SI{70}{\milli\second} narrowband metastable MOT is implemented. In this stage the modulation is turned off, the detuning is set to \SI{-150}{\kilo\hertz}, and the power is reduced to \SI{0.5}{\micro\watt}. 
Once the atoms have been loaded into the lattice, state preparation into $5s^2\,^1\mathrm{S}_0~M_F = \pm 9/2$ is carried out in a uniform bias magnetic field of \SI{64}{\micro\tesla} by implementing a \SI{15}{\milli\second} Doppler cooling stage at \SI{2.92}{\micro\meter} using the same large \SI{0.5}{\micro\watt} beam as used for the narrowband MOT, but red-detuned by \SI{100}{\kilo\hertz} from the cycling $5s5p\,^3\mathrm{P}_2~F = 13/2, M_F = 13/2 \rightarrow 5s5d\,^3\mathrm{D}_3~F = 15/2, M_F = 15/2$ transition. The atoms are then repumped to the ground state in \SI{5}{\milli\second} using beams at \SI{497}{\nano\meter} and \SI{679}{\nano\meter}, leaving 70\% of the atoms in the $5s^2\,^1\mathrm{S}_0$ $M_F = + 9/2$ state. To flip the atoms into $M_F = -9/2$, the field can be non-adiabatically switched to the opposite sign within \SI{100}{\micro\second}, and then adiabatically rotated back to its original direction in \SI{40}{\milli\second}. Finally, the atoms are selected by two filtering steps: first, the hot atoms are spilled by ramping the lattice down to a \SI{5}{\micro\kelvin} depth in \SI{20}{\milli\second}, holding for \SI{20}{\milli\second}, then ramping back up to the desired depth for spectroscopy. Second, the spin polarization purity is improved using a \SI{30}{\milli\second} Rabi $\pi$-pulse resonant with the clock transition for the desired spin state, followed by a \SI{2}{\milli\second} pulse at \SI{461}{\nano\meter} to clear out the atoms remaining in the ground state.
 With a typical cooling and state preparation cycle time of \SI{500}{\milli\second}, the system is capable of trapping \num{2e4} spin-polarized atoms at \SI{3}{\micro\kelvin} in the lattice.

\subsection*{The Lattice Enhancement Cavities}
\label{sec:lattices}

The cavity finesse of 6,000 and lattice waist of \SI{100}{\micro\meter} enables the system to generate a trap depth of \SI{30}{\micro\kelvin} with only \SI{1}{\milli\watt} of \SI{813}{\nano\meter} cavity-coupled power, more than sufficient for operating the clock. This large enhancement factor has advantages: it is compatible with less expensive low-power laser sources and it helps reduce amplified spontaneous emission incident on the atoms, which could otherwise be a problematic source of systematic uncertainty in the clock transition frequency. However, with a cavity length of \SI{35.9}{\milli\meter}, the high finesse also has the drawback of imposing a cavity linewidth of just \SI{700}{\kilo\hertz}, which exacerbates the technical problem of frequency-modulation to amplitude-modulation (FM to AM) conversion whereby fluctuations in the laser detuning from the cavity resonance will modulate the intracavity circulating power. The FM to AM conversion can be especially problematic if it causes amplitude noise at harmonics of the trap's motional frequencies, since this results in parametric heating of the atomic sample \cite{Savard1997}. Given that the typical trapping frequencies are >\SI{50}{\kilo\hertz}, high bandwidth control loops are needed to stabilise the laser-cavity detuning so as to avoid such heating.  

\begin{figure}[tb]
    \centering
    \includegraphics[width=.75\columnwidth]
    {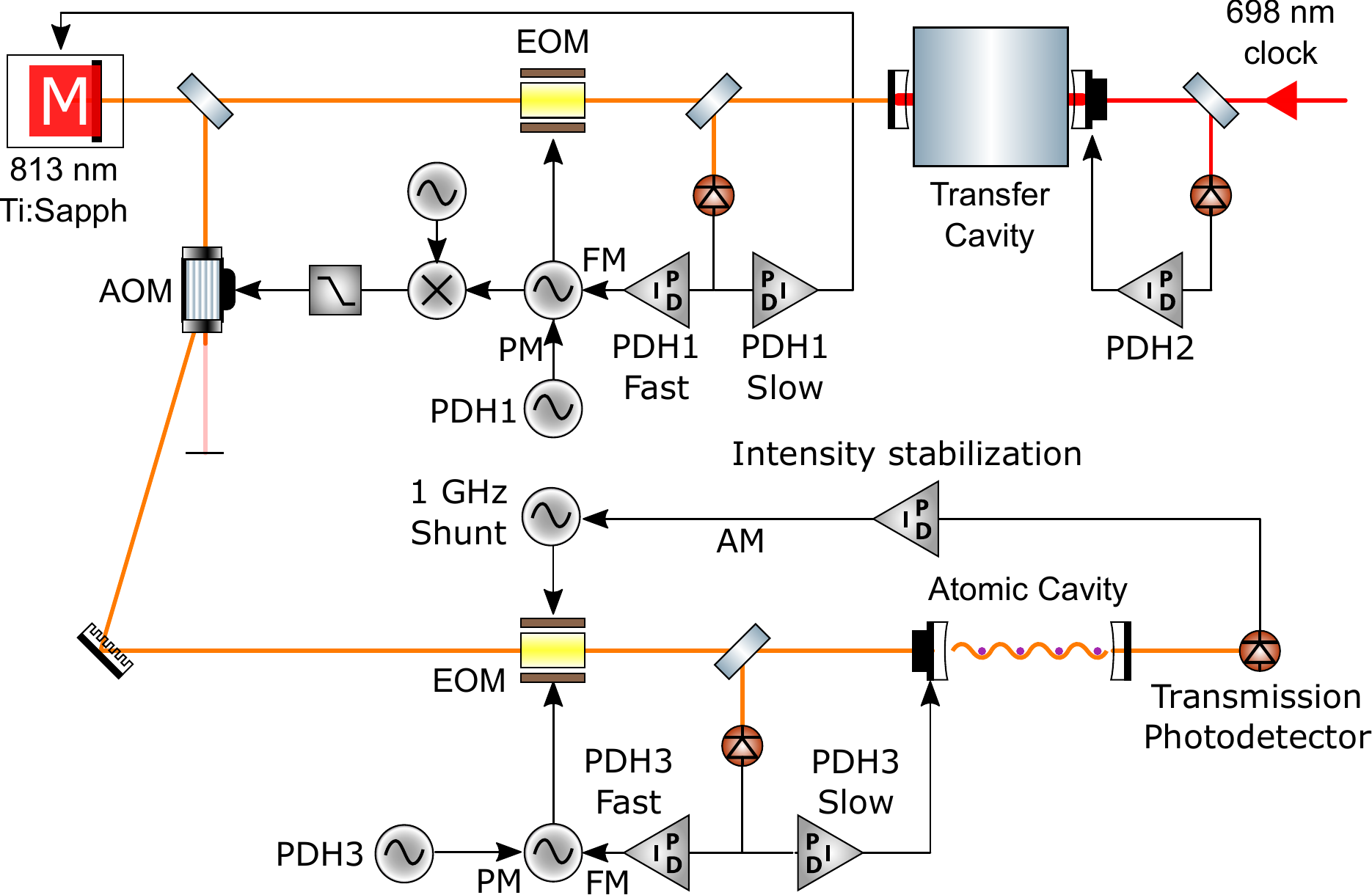}
    \caption{\label{fig:heating_schematic} Schematic showing the lattice laser and cavity stabilization scheme. For simplicity we only show one of the two atomic cavities. \textit{PDH1:} The 813 nm laser frequency is locked to a transfer cavity using the Pound-Drever-Hall (PDH) method. A fast feedback loop actuates on the 1\textsuperscript{st}-order sideband of a waveguide electro-optic modulator (EOM) in an electronic sideband configuration \cite{Thorpe2008}, and these fast frequency corrections are fed forward onto an acousto-optic modulator (AOM) to minimize frequency noise on the light sent to the atoms. A slow feedback loop then steers the \SI{813}{\nano\meter} laser to eliminate drift in the EOM sideband frequency. \textit{PDH2:} The length of the transfer cavity is locked to the \SI{698}{\nano\meter} clock laser to ensure that the \SI{813}{\nano\meter} laser is maintained at the magic wavelength with a long-term instability well below \SI{50}{\kilo\hertz}. \textit{PDH3:} In a similar arrangement to PDH1, a fast feedback loop couples the 1\textsuperscript{st}- order sideband of a waveguide EOM into the atomic cavity, and then a slow feedback loop steers the cavity length to eliminate drift in the EOM drive frequency. \textit{Intensity stabilization:} Finally, the transmission through the cavity is actively stabilized by shunting a variable amount of optical power into sidebands at $\pm\SI{1}{\giga\hertz}$ which are rejected from the cavity.}
\end{figure}

In our lattice-cavity system, the FM to AM conversion is minimized using the frequency stabilization scheme shown in figure \ref{fig:heating_schematic}. The system is designed both to ensure that the lattice is kept stable at the magic wavelength, and to lock the relative laser-cavity detuning tightly at the sub-\SI{10}{\kilo\hertz} level. Multiple control loops are needed in order to handle the free-running length instability of the atomic cavity: both in-vacuum cavities vibrate at a mechanical resonance frequency of \SI{9}{\kilo\hertz}, causing deviations of around \SI{100}{\kilo\hertz} on the \SI{813}{\nano\meter} mode which make it difficult to achieve a sufficiently tight cavity-laser lock using the piezoelectric actuators alone. Therefore, to reach sufficient loop gain we had to implement an additional high-bandwidth PDH lock of an \SI{813}{\nano\meter} sideband to the cavity---see figure \ref{fig:heating_schematic} for details.

\begin{figure}[tb]
    \centering
    \includegraphics[width=.99\columnwidth]{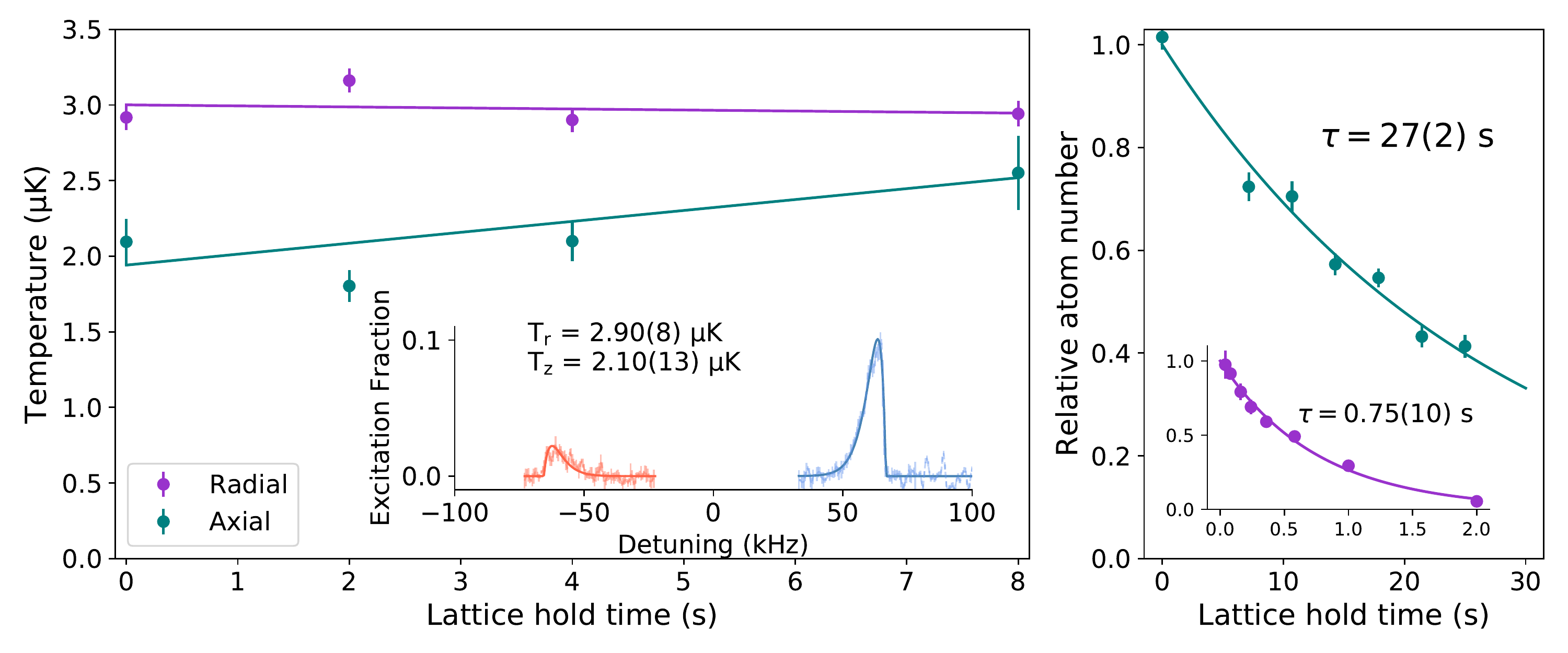}
    \caption{\label{fig:heating_data}\textit{Left:} Measured atomic temperature against hold time in the 1D optical lattice, with linear fits to the data. \textit{Inset:} Example sideband scan data used to infer the atomic temperature using the fitting function derived in \cite{Blatt2009}. \textit{Right:} Exponential decay of lattice-trapped atom number, limited by collisions with background gases. \textit{Inset:} Significantly increased decay rate of lattice-trapped atom number using only the slow piezo lock of the cavity length to the lattice laser, without the high-bandwidth lock of the laser to the atomic cavity (see figure \ref{fig:heating_schematic} for an explanation of these locks), showing the impact FM to AM conversion has on heating.}
\end{figure}

We used two methods to characterize the residual parametric heating rate in the cavity-enhanced lattice. The first method is to measure the lattice trap lifetime, and to compare it against the vacuum-limited lifetime of \SI{30}{\second} measured using magnetically trapped atoms in the 5s5p~$^{3}$P$_2$ state. As shown in the right panel of figure \ref{fig:heating_data}, the lattice lifetime is \SI{0.75}{\second} when only the slow PDH loops are engaged, but it increases to \SI{27}{\second} when the high-bandwidth PDH loops are added, indicating that parametric heating from FM to AM conversion can cause significant of atom loss. Whether or not the high-bandwidth PDH loops are engaged, the intensity stabilisation loop is observed to make no significant difference to the lifetime. This is most likely the result of the loop bandwidth being insufficient to significantly reduce noise above \SI{100}{\kilo\hertz}, which is required to suppress parametric heating.

To characterize the residual heating rate after all loops are engaged, we apply a second and more direct method: the atoms are held in the lattice for a fixed duration before measuring their temperature using resolved sideband spectroscopy of the clock transition as shown in the left panel of figure \ref{fig:heating_data} \cite{Bergquist1987}. Measuring the shape and amplitude of the sidebands yields information about the relative atomic populations in different motional states, which can then be used to infer the mean temperature \cite{Blatt2009}. For the axial direction, we measure a heating rate of less than 0.1~quanta/s, while in the transverse direction we do not resolve any change in temperature. 

\begin{figure}[tb]
    \centering
    \includegraphics[width=.75\columnwidth]{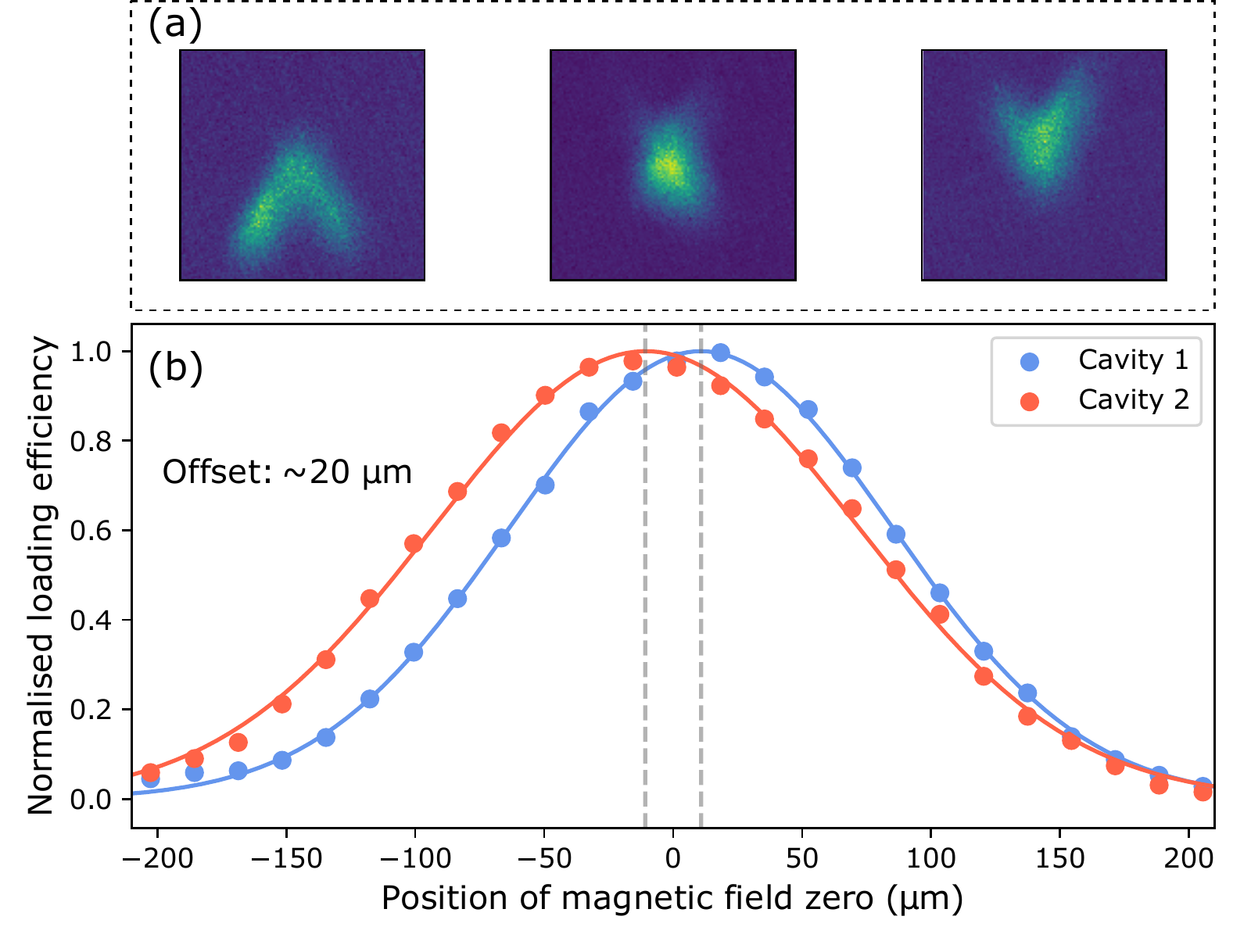} \caption{\label{fig:2dLattice} Finding the crossing point of the two cavities and measuring their relative overlap is a two step process. First, the position of the MOT is translated along the spectroscopy cavity by moving the magnetic field zero to find its intersection point with the other cavity. This process is shown in (a). Next a similar process is performed, but the cloud is now moved along a line perpendicular to both lattices, i.e. normal to the image plane shown in (a), to measure the position of each trap center. From Gaussian fits to the position dependent lattice loading efficiencies (shown in (b)), the relative offset between the two modes is estimated to be \SI{20}{\micro\meter}.}
\end{figure}

As outlined previously, the overlap of the two cavity waists was set outside of vacuum by tilting one of the mirrors in the non-spectroscopy cavity using a precision-polished dowel pin inserted beneath a flexure mount; details regarding this procedure are outlined in the Methods section. To verify that the cavities remained aligned after being placed under vacuum and undergoing bakeout at \SI{125}{\degreeCelsius}, we measured the position-dependent lattice loading efficiency as the second-stage MOT was translated across the two lattices. The results of this procedure are shown in figure \ref{fig:2dLattice}. We observed a final offset between the two in-vacuum cavities of approximately \SI{20}{\micro\meter}, comfortably less than the mode waist radius of \SI{100}{\micro\meter}, enabling the realization of a 2D lattice trap.

\begin{figure}[tb]
    \centering
    \includegraphics[width=.75\columnwidth]{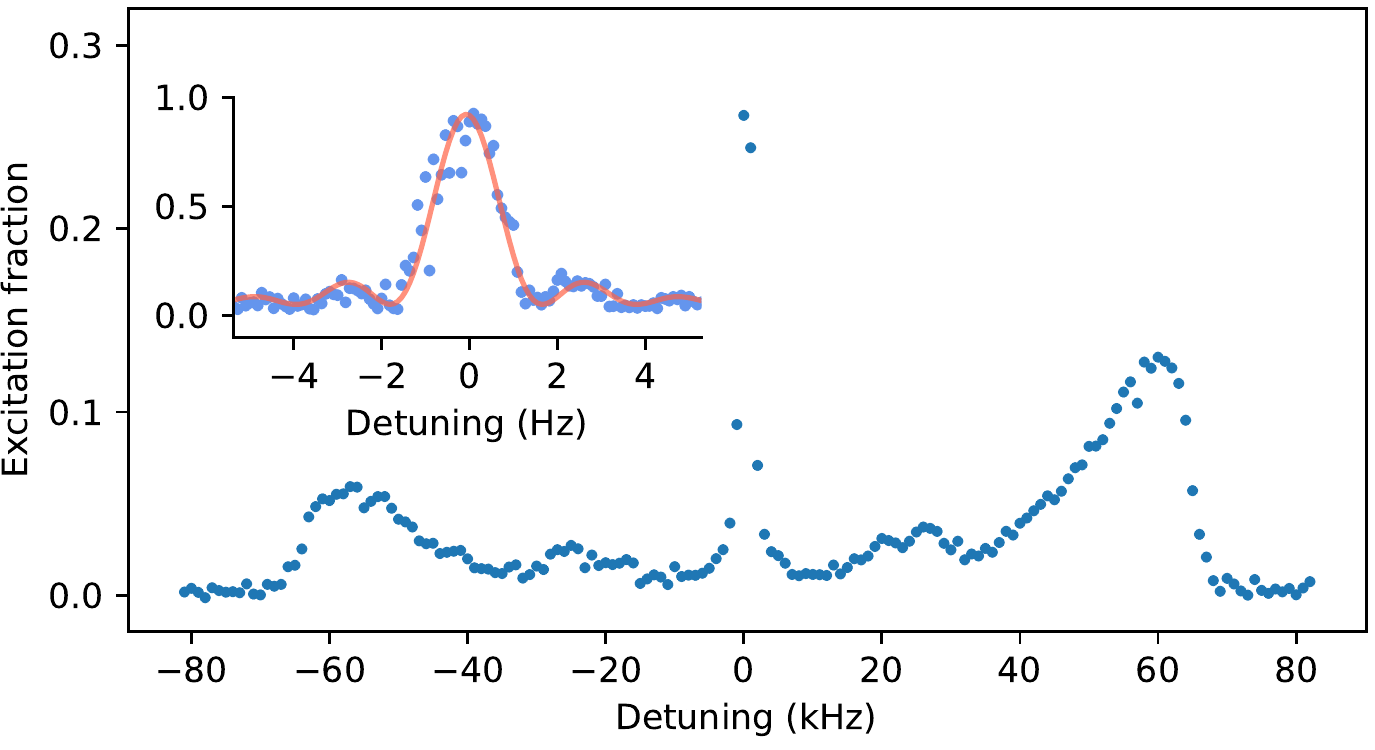} \caption{\label{fig:clock_scans} \textit{Main figure:} Average of four scans over motional sidebands of atoms trapped in the 2D lattice. The two distinct motional trapping frequencies at 27 and \SI{61}{\kilo\hertz} are clearly visible.  \textit{Inset:} Single high-resolution scan over the clock transition using a \SI{500}{\milli\second} Rabi $\pi$ pulse on a spin-polarized atomic sample.}
\end{figure}

To verify that the in-vacuum cavities can be used to realize an OLC, we carried out spectroscopy on the clock transition with the results shown in figure \ref{fig:clock_scans}. The \SI{698}{\nano\meter} clock transition in \textsuperscript{87}Sr is highly forbidden with a natural linewidth of around \SI{1}{\milli\hertz} \cite{Dorscher2018}, so we applied the electron shelving technique to detect the atomic excitation fraction after applying a clock spectroscopy pulse \cite{Dehmelt1982}. To resolve the red and blue motional sidebands in the 2D lattice, we excited the clock transition using a \SI{20}{\milli\second} probe pulse. The probe beam was aligned with the spectroscopy cavity mode and has a peak intensity of \SI{150}{\milli\watt\per\centi\meter^2}. The two separate motional frequencies of \SI{27}{\kilo\hertz} and \SI{61}{\kilo\hertz} indicate tight confinement in two directions, with implied trap depths of \SI{12.4}{\micro\kelvin} and $\SI{3.9}{\micro\kelvin}$ along the spectroscopy and the non-spectroscopy cavities, respectively. Finally, we took a high-resolution scan over the clock transition by reducing the probe intensity in order to drive a \SI{500}{\milli\second} Rabi $\pi$-pulse. Under these conditions we observed a transition contrast of 0.9 and a full width half maximum of \SI{1.7}{\hertz}, representing a $Q$ factor of \num{2.5e14} and providing a good basis to realize a stable OLC.

\subsection*{Systematic shifts in the optical lattice clock}
\label{sec:clock_compatibility}

Previously, it has been observed that the charging of dielectric surfaces near the atoms can result in DC Stark shifts at the $1\times 10^{-13}$ level \cite{Lodewyck2012}---more than four orders of magnitude larger than the target inaccuracy of our OLC. To verify that the accuracy of our system is not compromised by such effects, we followed the standard procedure\cite{Matveev2011}. First, voltages were applied to the in-vacuum electrodes to induce a Stark shift of the clock transition, which was then spectroscopically measured relative to an independent Sr frequency standard. The process was repeated, but with the field polarity reversed. If the difference in the two measured shifts is zero, the quadratic nature of the perturbation implies that there is no residual electric field component along the direction of the applied field. We repeated this measurement process along three orthogonal directions to characterize fully any residual background field, observing nearly symmetrical frequency shifts of a few \si{\hertz} from the applied electric fields along each axis. To avoid inadvertently charging dielectric surfaces with the applied voltages, the electrode voltages were reversed during part of the cooling stage such that the integrated electric field over each experimental cycle was zero. For a measurement time of less than one hour, we infer a total fractional DC Stark shift from the residual background electric field of $(-1.2 \pm 0.8)\times 10^{-18}$.

Collisions with background gas contaminants inside the vacuum chamber also contribute a systematic shift to the clock transition. Given the variety of in-vacuum materials in the pyramid MOT assembly, the possibility of poor vacuum quality was an important concern during the design stage. As a precautionary measure, large non-evaporable getter pumps with a total pumping speed of \SI[per-mode=symbol]{500}{\litre\per\second} for hydrogen were installed behind the pyramid MOT with an estimated conductance to the atom trapping region of \SI[per-mode=symbol]{300}{\litre\per\second}. To estimate an upper bound on the background gas pressure achieved in this system, the lifetime of atoms magnetically trapped in the $^3$P$_2$ state was measured \cite{Nagel2003}. The trap lifetime of \SI{30}{\second} indicates a pressure of approximately \SI{1e-10}{\milli\bar} \cite{Nagel2003}. By applying an appropriate model\cite{Gibble2013}, and assuming a background dominated by hydrogen (or by other gases with similar $C_6$ coefficients \cite{Mitroy2010a}), the associated collisional frequency shift is estimated to be \num{-5e-19}. Because the shift model is yet to be experimentally verified, the uncertainty is also taken to be \num{5e-19}.

For typical room-temperature OLCs using strontium or ytterbium, blackbody radiation (BBR) induces the largest systematic shift. In the most accurate systems, considerable effort has often been undertaken to minimize the BBR uncertainty, for example using in-vacuum temperature sensors \cite{Nicholson2015} or a well-characterized copper enclosure nested inside the vacuum chamber \cite{Beloy2014, McGrew2018}. In principle the pyramid MOT platform in this report would be compatible with these advanced methods, but we have opted instead for the simpler approach of affixing eight calibrated Pt100 temperature sensors to the outside of the vacuum chamber. The BBR temperature was then modeled as following a square probability distribution between the maximum and the minimum sensor readings, with a resulting temperature uncertainty given by $(T_\mathrm{max}-T_\mathrm{min})/\sqrt{12}$ \cite{Falke2014}. The temperature homogeneity in our system was aided by the compactness of the pyramid MOT assembly, which allowed the use of a relatively small titanium vacuum chamber with an outer radius of \SI{39}{\milli\meter}. As well as improving the thermal conductance between different sides of the vacuum chamber, the small chamber radius also reduces the required power for MOT coils resulting in significantly less heat dissipation. We have taken additional precautions by water cooling the MOT coils and separating them from the chamber with an air gap. The resulting peak-to-peak temperature inhomogeneity across the chamber when operating a clock sequence was \SI{200}{\milli\kelvin}, corresponding to a systematic fractional frequency uncertainty of \num{4e-18} for the BBR-induced AC Stark shift. 

A complete evaluation of the systematic frequency shifts in the OLC is currently underway. However, the data presented here indicates that the pyramid MOT and cavity assembly are compatible with a systematic uncertainty in the low \num{e-18} range.

\section*{Discussion}

\label{sec:conclusion}

We have realized a cold atom platform containing a pyramid MOT and two optical cavities, and we have demonstrated that it can prepare atoms in a suitable environment for a strontium optical lattice clock. Among other applications, the platform provides a promising route to realizing a compact, robust, transportable OLC with systematic uncertainty at the \num{e-18} level.

However, we recommend a few design modifications that could extend the capabilities of the system. Firstly, the silver-coated glass prisms used for the pyramid MOT could be replaced by all-metal substrates. This would allow the use of screws and alignment pins rather than epoxies to affix the prisms to the baseplate, and would also serve to eliminate a potential source of unwanted stray DC electric fields. Secondly, passive vibration-damping elements could be incorporated into the mounting structure of the vacuum chamber so that excitation of the mechanical resonances of the optical cavities could be substantially reduced. This simple step could obviate the need for the rather elaborate frequency stabilization system discussed above. Finally, a third optical cavity could be incorporated in the same platform, with mode overlap ensured using the flexure-based alignment methods, in order to realize a robust 3D optical lattice clock with either bosons \cite{Akatsuka2010} or fermions \cite{Campbell2017}.

Future work with the cold atom platform will focus on a full characterization and comparison with our first-generation strontium lattice clock \cite{Hill2016} and on techniques of cavity-enhanced, non-destructive measurement with the aim of realizing a spin-squeezed optical clock.

\bibliography{refs}

\section*{Methods}
\subsection*{Pyramid MOT construction}
The six prisms that reflect the beams needed for radial confinement in the MOT measure \SI{10}{\milli\meter} in width with a height of \SI{12.7}{\milli\meter}. The prism mirrors are glued on to a titanium baseplate using low out-gassing, UHV compatible epoxy (EPO-TEK H77). A hexagonal hole, with an inradius of \SI{4.3}{\milli\meter}, remains at the centre of the reflector which requires that an additional right-angle prism be placed underneath the baseplate to retro-reflect the central part of the MOT beam. The retro-reflecting prism is made from calcium fluoride (CaF$_2$), and its apex is machined to a precise right-angle without beveling so that aberrations in the reflected beam are minimized. The CaF$_2$ prism is glued on to stress-relieving flexure bars milled into the titanium baseplate. The flexure mounts mitigate thermal induced stress which is necessary to prevent cracking during UHV bakeout. CaF$_2$ was chosen primarily because of its transparency in the mid-infrared, making it compatible with realizing a second stage MOT operating on the $5s5p~^3\mathrm{P}_2$ to $5s4d~^3\mathrm{D}_3$ transition at \SI{2.92}{\micro\meter} (see figure \ref{fig:levels}). However, all the pyramid MOT optics are also compatible with the more conventional route to microkelvin temperatures with strontium; cooling on the $5s^2~^1\mathrm{S}_0$ to $5s5d~^3\mathrm{P}_1$ transition at \SI{689}{\nano\meter} \cite{Katori1999, Mukaiyama2003}. Another important feature of CaF$_2$ is the relatively low refractive index $n \approx 1.43$, which leads to near-critical total internal reflection at a 45 degree angle of incidence, thus minimizing polarization-dependent phase shifts upon retro-reflection of the MOT beam. Alternative optical materials with $n > 1.45$ would create a differential phase shift of more than 50 degrees between the $s$- and $p$-polarizations upon retro-reflection, severely undermining the MOT confinement force.

\subsection*{Enhancement cavity construction and alignment}

The cavity mirrors are glued on to the titanium baseplate and aligned such that the lattice beams fit through the gaps between the MOT prism mirrors. In order to ensure that each pair of mirrors is aligned sufficiently well to support cavity modes, they are clamped from above in a straight V-groove while the epoxy is cured. Once the cavity mirrors are secured to the baseplate, the overlap between the waists is set by tilting the cavity mirror using the flexure mount. The displacement between the two lattices was found by extinguishing the cavity modes by positioning a hypodermic needle into each beam using a translation stage. To adjust the tilt angle, we precision machined several dowel pin with diameters varying in steps of \SI{10}{\micro\meter}. This resolution corresponds to a minimum change in the cavity waist position of \SI{45}{\micro\meter} and resulted in a measured 2D cavity mode displacement of approximately \SI{5}{\micro\meter} upon initial setup for the optimum dowel pin diameter. 

\section*{Acknowledgements}

This work was financially supported by the UK Department for Business, Energy and Industrial Strategy as part of the National Measurement System Programme; and by the European Metrology Programme for Innovation and Research (EMPIR) project 15SIB03-OC18. This project has received funding from the EMPIR programme co-financed by the Participating States and from the European Union’s Horizon 2020 research and innovation programme. 

\section*{Author contributions statement}
W.B., R.H. and I.R.H designed the system. W.B. and  R.H. built the apparatus, conducted experiments and wrote the manuscript. W.B., R.H. and A.V. analyzed the data and constructed the figures. A.S. and H.S.M built and operated the measurement infrastructure needed for the stability transfer of the ultra-stable \SI{1064}{\nano\meter} laser (built and characterised by M.S.) to the clock wavelength. P.E.G.B. and P. G. supervised this work. All authors reviewed and commented on the manuscript. 

\section*{Competing Interests}
The authors declare no competing interests.

\end{document}